\documentstyle[11pt,draft]{article}
\newcommand{\be}{\begin{equation}}
\newcommand{\ee}{\end{equation}}

\newcommand{\real}{\mbox{{\rm I\hspace{-2truemm} R}}}

\newcommand{\di}{\mbox{{\rm dim}}}
\newcommand{\diag}{\mbox{{\rm diag}}}
\newcommand{\id}{\mbox{{\rm 1\hskip-1truemm I}}}

\newcommand{\Tr}{\mbox{{\rm Tr}}}

\title{The Poincar\'e coset models ISO(d-1,1)/$\real^n$ and T-duality}
\author{Roberto Casadio\thanks{
e-mail: casadio@bo.infn.it}
\ and Benjamin Harms\thanks{e-mail: bharms@ua1vm.ua.edu}\\
 \\
{\em
Department of Physics and Astronomy, The University of Alabama}\\
{\em Box 870324, Tuscaloosa, AL 35487-0324}}
\begin{document}
\begin{titlepage}
\pagestyle{empty}
\maketitle
\begin{abstract}
We generalize a family of Lagrangians with values in the
Poincar\'e group $ISO(d-1,1)$, which contain the description of
spinning strings in flat $(d-1)+1$ dimensions, by including
symmetric terms in the world-sheet coordinates.
Then, by promoting a subgroup $H\sim\real^n$, $n\le d$,
which acts invariantly from the left on the element of
$ISO(d-1,1)$, to a gauge symmetry of the action, we obtain
a family of $\sigma$-models.
They describe bosonic strings moving in (generally) curved,
and in some cases degenerate, space-times with an axion field.
Further, the space-times of the effective theory
admit in general T-dual geometries.
We give explicit results for two non degenerate cases.
\end{abstract}
Pacs: 04.20.DW, 11.25.-w, 11.30.Cp
\par\noindent
Keywords:\par
axion field\par
coset construction\par
curved space-times\par
Poincar\'e invariance\par
sigma models\par
T-duality\par
\rightline{UAHEP973}
\end{titlepage}
%
\pagestyle{plain}
\raggedbottom
\setcounter{page}{1}
\section{Introduction}
Recently several string actions which
naturally describe curved space-times with singularities
have been obtained from WZWN models.
Following the coset construction \cite{witten,HH,bars},
in Ref.~\cite{ch} we analyzed a particular WZWN action
in the Poincar\'e $ISO(2,1)$ group.
\par
In a recent paper \cite{stern} a very general family of Lagrangians
in the Poincar\'e group $ISO(d-1,1)$ was studied and shown to describe
diverse closed, bosonized, spinning strings in $(d-1)+1$-dimensional
Minkowski space-time depending on the values of the constants which
parameterize the family.
In this paper, we start from the family of actions cited above and add
further contributions amounting to terms which are symmetric in the
world-sheet indices.
Then we apply a gauging procedure which generalizes the one used in
Ref.~\cite{ch}:
we raise a subgroup $H$ to a local symmetry of the action, where
$H$ turns out to be necessarily isomorphic to $\real^n$, $n\le d$,
because of the prescription that the gauge field belongs to the algebra
of $H$ itself, and show that the gauged action generates a family
of effective actions for $\sigma$-models with $N=d\,(d+1)/2-\di(H)$
degrees of freedom.
The latter can be viewed as effective theories describing the
dynamics of a bosonic string moving in a (generally) curved
background which can also contain an axion field.
\par
Further, since both the metric and axion fields are independent
of at least $d-\di(H)$ out of $N$ degrees of freedom,
it is easy to prove that the usual T-duality
considerations apply \cite{giveon}.
One can include a dilaton field at a higher order in the loop
expansion and build dual spaces.
\par
The main idea behind the procedure we use is actually quite simple,
and it is worth displaying the way it works on a {\em toy model\/}
to show its main features.
Consider the following 2-dimensional action,
\be
S(x^1,x^2)=
{1\over 2}\,\int dt\,\left[(\partial_t x^1)^2+(\partial_t  x^2)^2\right]
\ ,
\ee
and gauge one of the coordinates, {\em e.g.\/} $x^2$,
by the minimal coupling prescription,
$\partial_t  x^2\to \partial_t  x^2+A^2$, where $A^2$ is a gauge field,
\be
S_g(x^1,x^2,A^2)=
S+{1\over 2}\,\int dt\,A^2\left(A^2+2\,\partial_t  x^2\right)
\ .
\ee
Since $S_g$ is quadratic in $A^2$, one can define an effective
action by integrating out $A^2$ in the path integral,
\be
\int [dx^1]\,[dx^2]\,[d A^2]\,e^{-S_g(x^1,x^2,A^2)}\equiv
\int [dx^1]\,e^{-S_{eff}(x^1)}
\ .
\ee
This is equivalent to solving the equation of motion for $A^2$,
$\delta_{A^2} S_g=0$, and substituting back the result into
$S_g$.
Thus one obtains
\be
S_{eff}(x^1)={1\over 2}\,\int dt\,(\partial_t  x^1)^2
\ ,
\ee
which is a trivial result and equals the one which we get
by assuming $A^2$ is a pure gauge, $A^2=-\partial_t  x^2$.
But suppose we now perform the following (canonical)
transformation,
\be
\left[\begin{array}{c}
x^1 \\
x^2
\end{array}\right]
\equiv
\left[\begin{array}{cc}
\theta_{11} & \theta_{12} \\
\theta_{21} & \theta_{22}
\end{array}\right]
\,
\left[\begin{array}{c}
x \\
\tilde x
\end{array}\right]
\ ,
\label{can}
\ee
with $\theta_{11}\,\theta_{22}-\theta_{12}\,\theta_{21}\not=0$,
then we gauge {\em e.g.\/} $\tilde x$ introducing a gauge field
$\tilde A$ and repeat the process above.
This time we obtain
\be
S_{eff}(x)={1\over 2}\,\int dt\,
{(\theta_{11}\,\theta_{22}-\theta_{12}\,\theta_{21})^2\over
\theta_{12}^2+\theta_{22}^2}\,(\partial_t x)^2
\ ,
\ee
which is different from the result one would get by setting
$\tilde A=-\partial_t \tilde x$, namely
\be
S(x,\tilde x=const)={1\over 2}\,\int dt\,(\theta_{11}^2+\theta_{21}^2)\,
(\partial_t x)^2
\ .
\ee
We can rephrase this conclusion by saying that the canonical
transformation in Eq.~(\ref{can}) introduces {\em cross terms\/}
of the form $\partial_t x\,\partial_t \tilde x$ in the action
and these in turn generate the following mapping:
\be
(\theta_{11}^2+\theta_{21}^2)\to
{(\theta_{11}\,\theta_{22}-\theta_{12}\,\theta_{21})^2\over
\theta_{12}^2+\theta_{22}^2}
\ ,
\label{toy_map}
\ee
which becomes trivial for
$\theta_{11}\,\theta_{12}+\theta_{21}\,\theta_{22}=0$
(this also rules out rotations).
\par
Of course, the previous model applies to quadratic actions only.
Whenever we encounter Lagrangians which are linear in the gauge
field we will revert to the pure gauge sector, as we did in
Ref.~\cite{ch}, and obtain degenerate metrics (apart from two
exceptional cases).
\par
This in brief describes both our coset construction
and compactification to get T-dual solutions.
However,
due to the high degree of generality of the model we start with,
we were not able to draw any explicit conclusions other than
formal mappings like the one shown in Eq.~(\ref{toy_map})
for the toy model above.
In particular, we cannot say much about the reduced space-time
in general, although we show that,
with particular choices of the parameters
involved and for $d=2,3$, it is actually possible to complete
the analysis.
\par
In the next Section we describe the ungauged model,
with particular attention to the derivation of the equations
of motion and their comparison with the models introduced in
Ref.~\cite{stern}.
In Section~\ref{gauged} we give the general formal treatment of
the action when one gauges a subgroup $H\sim\real^n$,
identify the whole set of its symmetries and introduce an effective
action in the form of a $\sigma$-model.
We obtain expressions for actions which can be
either quadratic or
linear in the gauge field, but in the latter case we show there
are only two cases with non degenerate metrics.
In Section~\ref{T-d} we describe the T-dual procedure as applied
to our model and show the properties of our model which are related
to its multiple isometries.
We then study the properties of the effective action under T-duality
transformations.
In Section~\ref{d=2} we specialize to the simplest, 2-dimensional,
non degenerate case and perform an explicit analysis to find one
effective background and one of its T-duals.
In Section~\ref{iso21} we prove that the (non degenerate) metrics
for all the models with $d=3$ which are linear in the gauge field
reduce to the case already treated in Ref.~\cite{ch},
which we now revise.
\setcounter{equation}{0}
\section{The ungauged action}
\label{ungauged}
We recall here that the Poincar\'e group in $(d-1)+1$ space-time
dimensions, $ISO(d-1,1)$, is the semidirect product of the Lorentz
group $SO(d-1,1)$ with the space-time translation group
$T(d-1,1)\sim\real^{(d-1,1)}\sim\real^d$.
Therefore we write its elements $g$ using the notation
$g=\left(\Lambda,x\right)$, where $\Lambda\in SO(d-1,1)$ and
$x\in \real^d$.
\par
Given the map $g:\ {\cal M}\mapsto ISO(d-1,1)$ from the
2-dimensional manifold ${\cal M}$, parametrized by the
coordinates $\sigma^\alpha$ ($\sigma^0\equiv\tau$,
$\sigma^1\equiv\sigma$), to $ISO(d-1,1)$,
we consider the very general action given by
\be
S(\Lambda,x;K)=S_1+S_2+S_3
\ ,
\label{S}
\ee
where
\begin{eqnarray}
S_1&=&{1\over2}\,\int_{\cal M}\,d^2\sigma\,
(g^{\alpha\beta}+\epsilon^{\alpha\beta})\,
K_{ij}^{(1)}\,V^i_\alpha\,V^j_\beta
\nonumber \\
S_2&=&\int_{\cal M}\,d^2\sigma\,(g^{\alpha\beta}+\epsilon^{\alpha\beta})\,
K_{ijk}^{(2)}\,V^i_\alpha\,W^{jk}_\beta
\label{S123}  \\
S_3&=&{1\over8}\,\int_{\cal M}\,d^2\sigma\,
(g^{\alpha\beta}+\epsilon^{\alpha\beta})\,
K_{ijkl}^{(3)}\,W^{ij}_\alpha\,W^{kl}_\beta
\ .  \nonumber
\end{eqnarray}
Summation is assumed among upper and lower repeated latin indices
$i,j,\dots=0,\dots,d-1$ according to the usual Lorentzian scalar product
rule $A_i\,B^i\equiv A^i\,\eta_{ij}\,B^j$, where
$\eta_{ij}=(-,+,\dots,+)$
is the Minkowski tensor in $(d-1)+1$-dimensions.
\par
The Lagrangians in Eqs.~(\ref{S123}) contain two kinds of contribution:
the first one is proportional to the area element on ${\cal M}$,
$d^2\sigma\,\epsilon^{\alpha\beta}=d\sigma^\alpha\wedge d\sigma^\beta$,
with $\epsilon^{\alpha\beta}=-\epsilon^{\beta\alpha}$
($\epsilon^{\tau\sigma}=+1$) the Levi Civita symbol in two dimensions;
the second one is proportional to the constant symmetric 2-dimensional
matrix $g^{\alpha\beta}$ with $|\det g^{\alpha\beta}|=1$.
Since the constants $K$ are assumed to satisfy
\begin{eqnarray}
K^{(1)}_{ij}&=&-K^{(1)}_{ji}
\nonumber \\
K^{(2)}_{ijk}&=&-K^{(2)}_{ikj}
\nonumber \\
K^{(3)}_{ijkl}&=&-K^{(3)}_{jikl}=-K^{(3)}_{ijlk}=-K^{(3)}_{klij}
\ ,
\label{K_cond}
\end{eqnarray}
it turns out that the contributions proportional to $g^{\alpha\beta}$
drop out both of $S_1$ (because of the skewsymmetry of $K^{(1)}_{ij}$
in the indices $i,j$)
and $S_3$ (because of the skewsymmetry of $K^{(3)}_{ijkl}$
under the exchange of the pairs of indices $(i,j),(k,l)$)
and thus only $S_2$ contains it.
When $g^{\alpha\beta}\equiv 0$, the action $S(K)$ coincides with the
model introduced in Ref.~\cite{stern} and describes a different
kind of closed bosonized spinning string moving in $(d-1)+1$ Minkowski
space-time with coordinates $x^k$, $k=0,\dots,d-1$ depending
on the values of the constants $K$.
\par
The 1-forms $V^i$, $W^{ij}$, with components
\be
\begin{array}{l}
V^i_\alpha\equiv(\Lambda^{-1})^i_{\ r}\,\partial_\alpha x^r  \\
\\
W^{ij}_\alpha\equiv
(\Lambda^{-1})^i_{\ r}\,\partial_\alpha\Lambda^{rj}
\ ,
\end{array}
\label{gdg}
\ee
are obtained by projecting the (left invariant)
Maurer-Cartan form $g^{-1}\,dg$ on the basis of the Poincar\'e
algebra $iso(d-1,1)$.
Thus it immediately follows that the action $S$ in Eq.~(\ref{S})
is invariant under the left rigid action of the Poincar\'e group,
$g\to g'\,g$, $g'=(\theta,y)\in ISO(d-1,1)$,
\be
S(\theta\,\Lambda,\theta\,x+y;K)=S(\Lambda,x;K)
\ .
\ee
It is also invariant under the right rigid action,
$g\to g\,g'$, $g'=(\theta,y)\in ISO(d-1,1)$,
\be
S(\Lambda\,\theta,\Lambda\,y+x;K)=S(\Lambda,x;K')
\ ,
\ee
provided the constants
$K\equiv(K^{(1)}_{ij},K^{(2)}_{ijk},K^{(3)}_{ijkl})$
map to new values $K'$ according to an expression given in
Ref.~\cite{stern}.
The action $S_2$ alone, depending on the choice of $g^{\alpha\beta}$,
may actually be invariant under a {\em semi\/}-local transformation,
as we report in Section~\ref{iso21}.
\par
If we use the convention $\Lambda_i^{\ j}\equiv(\Lambda^{-1})_{\ i}^j$,
the action $S$ can be written more explicitely in terms of the
elements $\Lambda$ and $x$ as
\begin{eqnarray}
S_1&=&{1\over2}\,\int_{\cal M}\,d^2\sigma\,\epsilon^{\alpha\beta}\,
K_{ij}^{(1)}\,
\Lambda^{\ i}_r\,\Lambda^{\ j}_s\,
\partial_\alpha x^r\,\partial_\beta x^s
\nonumber  \\
S_2&=&\int_{\cal M}\,d^2\sigma\,(g^{\alpha\beta}+\epsilon^{\alpha\beta})\,
K_{ijk}^{(2)}\,
\Lambda^{\ i}_r\,\Lambda^{\ j}_s\,
\partial_\alpha x^r\,\partial_\beta \Lambda^{sk}    \\
S_3&=&{1\over 8}\,\int_{\cal M}\,d^2\sigma\,\epsilon^{\alpha\beta}\,
K_{ijkl}^{(3)}\,
\Lambda^{\ i}_r\,\Lambda^{\ k}_s\,
\partial_\alpha \Lambda^{rj}\,\partial_\beta \Lambda^{sl}
\nonumber
\ .
\end{eqnarray}
\par
The equations of motion $\delta_x S=0$, which follow from the
variation $x\to x+\delta x$, with $\delta x$ an infinitesimal
$(d-1)+1$ vector, amount to linear momentum conservation,
\be
\partial_\alpha{\cal P}^\alpha_i\equiv
\partial_\alpha\left(\epsilon^{\alpha\beta}\,{\cal P}_{\beta\,i}^{(1)}
+(g^{\alpha\beta}+\epsilon^{\alpha\beta})\,
{\cal P}_{\beta\,i}^{(2)}\right)
=0
\ ,
\label{dp=0}
\ee
where the only two linear momentum currents that are not identically
zero follow from $S_1$ and $S_2$ and are respectively given by
\be
\begin{array}{l}
{\cal P}_{i}^{(1)\alpha}=\epsilon^{\alpha\beta}\,
\Lambda^{\ r}_i\,K_{rs}^{(1)}\,V^s_\beta
\equiv {\cal V}^{\,\alpha\beta}_{is}\,\partial_\beta x^s \\
  \\
{\cal P}_{i}^{(2)\alpha\,}=
\left(g^{\alpha\beta}+\epsilon^{\alpha\beta}\right)\,
\Lambda^{\ r}_i\,K_{rst}^{(2)}\,
W^{st}_\beta\equiv
{\cal W}^{\,\alpha\beta}_{ist}\,\partial_\beta\Lambda^{st}
\ .
\end{array}
\label{T}
\ee
Upon integrating on a fixed $\tau$ slice of the world-sheet,
one finds that the conserved charges are given by
\be
P^{(1)}_i+P^{(2)}_i=\int d\sigma\ \left[{\cal P}^{(1)}_{\sigma\,i}
+(1+g^{\tau\sigma})\,{\cal P}^{(2)}_{\sigma\,i}
+g^{\tau\tau}\,{\cal P}^{(2)}_{\tau\,i}\right]
\ ,
\ee
where $\tau$ and $\sigma$ are the world sheet coordinates and use has
been made of the periodicity in $\sigma$ to discard boundary terms.
For the particular choice we will make in Section~\ref{iso21},
$g^{\alpha\beta}=\pm\eta^{\alpha\beta}=
\pm\diag(-1,1)$ (the Minkowski metric tensor on the world-sheet),
one finds that the conserved linear momentum following from $S_2$
coincides with the spatial integral of ${\cal P}^{(2)}_\pm$,
where
\be
\sigma^\pm\equiv\tau\pm\sigma
\label{light}
\ee
are light-cone coordinates on the world-sheet.
\par
Similarly, from the variation $\Lambda\to\Lambda+\delta\Lambda$,
$\delta\Lambda=\Lambda\,\rho$ and $\delta x=\rho\,x$,
with $\rho_{ij}=-\rho_{ji}$ an infinitesimal $so(d-1,1)$
matrix, the equations $\delta_\Lambda S=0$ lead to
angular momentum conservation
\be
\partial_\alpha{\cal J}^\alpha_{ij}\equiv
\partial_\alpha\left(\epsilon^{\alpha\beta}\,{\cal J}^{(1)}_{\beta\,ij}+
(g^{\alpha\beta}+\epsilon^{\alpha\beta})\,{\cal J}_{\beta\,ij}^{(2)}+
\epsilon^{\alpha\beta}\,{\cal J}_{\beta\,ij}^{(2)}\right)
=0
\ ,
\label{dj=0}
\ee
where the three angular momentum currents
following from $S_1$, $S_2$ and $S_3$ read
\begin{eqnarray}
{\cal J}_{ij}^{(1)\alpha}&=&{\cal L}_{ij}^{(1)\alpha}=
x_i\wedge{\cal P}_{j}^{(1)\alpha}
\nonumber \\
{\cal J}_{ij}^{(2)\alpha}&=&{\cal L}_{ij}^{(2)\alpha}
+{\cal S}_{ij}^{(2)\alpha}
\ ,\ \ \ \
\left\{\begin{array}{l}
{\cal L}_{ij}^{(2)\alpha}=x_i\wedge{\cal P}_{j}^{(2)\alpha}  \\
 \\
{\cal S}_{ij}^{(2)\alpha}=2\,{\cal W}_{rij}^{\alpha\beta}\,
\partial_{\beta}x^r
\end{array}\right.
\nonumber \\
{\cal J}_{ij}^{(3)\alpha}&=&{\cal S}_{ij}^{(3)\alpha}=
-{1\over2}\,\epsilon^{\alpha\beta}\,\Lambda_i^{\ r}\,\Lambda_j^{\ s}\,
K_{rstp}^{(3)}\,W_\beta^{tp}
\ .
\label{J}
\end{eqnarray}
It is thus clear that one obtains terms which can be interpreted as
non zero intrinsic angular momentum
(spin) ${\cal S}_{ij}$ without the use of Grassmann variables.
If one expands Eq.~(\ref{dj=0}) one finds that the conserved charges
are given by
\be
J^{(1)}_{ij}+J^{(2)}_{ij}+J^{(3)}_{ij}=
\int d\sigma\left[{\cal J}^{(1)}_{\sigma\,ij}
+(1+g^{\tau\sigma})\,{\cal J}^{(2)}_{\sigma\,ij}
+g^{\tau\tau}\, {\cal J}^{(2)}_{\tau\,ij}
+{\cal J}^{(3)}_{\sigma\,ij}\right]
\ ,
\ee
and, for $g^{\alpha\beta}=\pm\eta^{\alpha\beta}$, one obtains
$J^{(2)}=\int d\sigma{\cal J}^{(2)}_\mp$.
Again, this will be the case for the model studied in
Section~\ref{iso21}.
\par
To summarize, the difference between the models in Ref.~\cite{stern}
and ours is given by the contribution to linear and angular momentum
proportional to $g^{\alpha\beta}$ (see Eqs.~(\ref{T}), (\ref{J})).
\setcounter{equation}{0}
\section{The gauged action}
\label{gauged}
We can modify the action $S$ in Eq.~(\ref{S}) in such a way as to make
it invariant under the {\em local\/} left action of a subgroup $H$
of the whole Poincar\'e group,
\be
g\to h\,g=(\theta\,\Lambda, \theta\,x+y)
\ ,
\ee
with $h(\tau,\sigma)=(\theta,y)\in H$.
For this purpose we introduce a gauge connection
$A_\alpha(\tau,\sigma)=(\omega_\alpha,\xi_\alpha)$
and the corresponding covariant derivative
$D_\alpha g\equiv\partial_\alpha g+A_\alpha$.
\par
We require that $A_\alpha$ belongs to the algebra ${\cal H}$
of the group $H$ (so that it has as many components as the
elements of $H$ have). Since for every element $g\in ISO(d-1,1)$
one has
\be
D_\alpha(h\,g)\simeq\partial_\alpha g+\partial_\alpha(\delta h\,g)
+A_\alpha
\ ,
\ee
where $\delta h=(\delta\theta,\delta y)\in{\cal H}$, it follows that
$H$ must act invariantly from the left on the elements of $ISO(d-1,1)$,
that is
\be
\delta_{_L} g=\delta h\,g
=(\delta\theta\,\Lambda,\delta\theta\,x+\delta y)
\in {\cal H}\ ,
\ \ \ \forall\, g=(\Lambda,x)\in ISO(d-1,1)
\ .
\ee
The only possible non trivial choices for $H$ are then subgroups
of the translation group $\real^d$, that is
\be
\left\{\begin{array}{l}
\theta=0 \\
 \\
y\in \real^n\oplus\id_{d-n}\sim\real^n
\ ,
\end{array}\right.
\ee
with $n\le d$, $\id_{d-n}$ being the identity in $d-n$ dimensions,
for which $\delta_{_L} g=\delta h$, $\forall\, g\in ISO(d-1,1)$.
Thus one also has the following form for the gauge field
\be
\left\{\begin{array}{ll}
\omega_\alpha^k\equiv 0\ \ &\ \  k=0,\dots,d-1 \\
& \\
\xi_\alpha^k\equiv 0\ \ &\ \  k\not\in H
\ ,
\end{array}\right.
\label{xi}
\ee
where $k\not\in H$ is a shorthand notation for $x^k\not\in H$.
\par
Further, the gauge field must change under an infinitesimal
$ISO(d-1,1)$ transformation according to
\be
\left\{\begin{array}{ll}
\xi_\alpha^i\to \xi_\alpha^i-\partial_\alpha(\delta x^i)
\ \ &\ \ i\in H \\
& \\
\xi_\alpha^i\to \xi_\alpha^i=0
\ \ &\ \ i\not\in H
\ ,
\end{array}\right.
\label{cond}
\ee
where $\delta x$ is any allowed infinitesimal variation of $x\in\real^d$,
including $(d-1)+1$ Lorentz transformations.
\par
The gauged action then reads
\be
S_g(\Lambda,x,\xi;K)=S_{1g}+S_{2g}+S_{3g}
\ ,
\label{S_g}
\ee
with
\begin{eqnarray}
S_{1g}&=&{1\over 2}\,\int d^2\sigma\,{\cal V}^{\alpha\beta}_{rs}\,
\left(\partial_\alpha x+\xi_\alpha\right)^r\,
\left(\partial_\beta x+\xi_\beta\right)^s
\equiv S_1+\Delta S_1(\xi)
\nonumber       \\
S_{2g}&=&\int d^2\sigma\,{\cal W}^{\alpha\beta}_{rsk}\,
\left(\partial_\alpha x+\xi_\alpha\right)^r
\partial_\beta\Lambda^{sk}
\equiv S_2+\Delta S_2(\xi)  \label{S_123g}\\
S_{3g}&=&S_3
\ , \nonumber
\end{eqnarray}
with ${\cal V}$ and ${\cal W}$ defined in Eq.~(\ref{T}).
The new term
\be
\Delta S_1=\int d^2\sigma\,\sum\limits_{s\in H}\,
\left[{\cal P}^{(1)\alpha}_s\,\xi_\alpha^s+
{1\over 2}\,\sum\limits_{r\in H}\,
{\cal V}^{\,\alpha\beta}_{rs}\,\xi^r_\alpha\,\xi_\beta^s
\right]
\ ,
\label{ds1}
\ee
is bilinear in the gauge field $\xi$ while
\be
\Delta S_2=\int d^2\sigma\,\sum\limits_{s\in H}\,
{\cal P}^{(2)\alpha}_s\,\xi_\alpha^s
\ ,
\label{ds2}
\ee
is linear in $\xi$ ($\sum_{r,s,\dots\in H}$ means the indices $r,s,\dots$
are summed only over $r,s,\dots\in H$, while all other latin indices are
not restricted).
\subsection{Symmetries}
We now describe the symmetries of the new action.
To simplify the notation, we momentarily turn to the Euclidean case
$ISO(d)$ which is the semidirect product of the rotation group
$SO(d)$ with $\real^d$ (this can be achieved by complexifying
the time-like variable $x^0\mapsto i\,x^d$).
We then consider the following four subgroups:
\begin{itemize}
\item
the two groups of $n$- and $(d-n)$-dimensional translations,
with $n\le d$,
\be
\begin{array}{lcr}
H_n\equiv\real^n\oplus\id_{d-n}\ &\ {\rm and}\ &
H_{d-n}\equiv\id_n\oplus\real^{n-d}
\ ,
\end{array}
\ee
such that $\real^d\sim H_n\oplus H_{d-n}$,
\end{itemize}
and
\begin{itemize}
\item
the following two rotation subgroups of the whole rotation group:
\be
\begin{array}{lcr}
R_n\equiv SO(n)\oplus\id_{d-n}\ &\ {\rm and}\ &
R_{d-n}\equiv \id_n\oplus SO(d-n)
\ .
\end{array}
\ee
\end{itemize}
We also write the Euclidean connection $\bar \xi$ as a $d$-dimensional
vector $\bar\xi\equiv(\xi^k,\phi^\mu)$, $k=1,\dots,n$,
$\mu=n+1,\dots,d$, so that we separate its components into $\xi\in H_n$
and $\phi\in H_{d-n}$.
\par
The ungauged action obtained by Euclideanizing $S$ in Eq.~(\ref{S})
is invariant under the left rigid action of $ISO(d)$.
The Euclideanized action $S_g$ obtained by gauging the group $H=H_n$
becomes invariant under the left local action of $H_n$, and it
is still invariant under the left rigid action of $H_{d-n}$.
However it is no longer invariant under the left rigid action of the
whole $SO(d)$ group because $\Delta S_1(\xi)$ in Eq.~(\ref{ds1})
and $\Delta S_2(\xi)$ in Eq.~(\ref{ds2}) are not.
In fact, in order to preserve the invariance in $\Delta S_1$
and $\Delta S_2$, the gauge field $\bar\xi$ must transform
according to the (Euclidean version) of the first constraint
in Eq.~(\ref{cond}),
\be
\bar\xi_\alpha^i\to\bar\xi_\alpha^i-\theta^i_{\ l}\,\partial_\alpha x^l
\ ,\ \ \ \ \ \forall\, i=1,\dots,d
\ ,
\ee
but this would mix $\xi$ components with $\phi$ components of the
connection for a general rotation $\theta\in SO(d)$.
At the same time, according to Eq.~(\ref{xi}) or, equivalently, the
second constraint in Eq.~(\ref{cond}), it must be possible to
set $\phi$ to zero (or $\bar\xi^i=0$, $\forall\, i=n+1,\dots,d$),
since these components correspond to the group $H_{d-n}$ that we are not
gauging.
This implies that the only rigid rotations that leave $S_g$ left invariant
are the ones which do not mix $\xi$ with $\phi$ and thus belong
to $R_n$ or $R_{d-n}$.
To summarize:
\be
S_g(\theta\,\Lambda,\theta\,x+y,\xi;K)=S_g(\Lambda,x,\xi;K)
\Leftrightarrow\left\{
\begin{array}{l}
y(\tau,\sigma)\in H_n\ \vee\ y\in H_{d-n} \\
\\
\theta\in R_n\ \vee \ \theta\in R_{d-n}
\ .
\end{array}\right.
\ee
\par
The same argument above applied to right rigid transformations
would require
\be
\bar\xi_\alpha\to\bar\xi_\alpha-(\partial_\alpha\Lambda)\,y
\ ,\ \ \ \ \ \forall\,\Lambda\in SO(d)
\ ,
\ee
which necessarily mixes $\xi$ and
$\phi$ components if $y\not\equiv 0$.
This singles out the whole $SO(d)$ subgroup, so that
\be
S_g(\Lambda\,\theta,\Lambda\,y+x,\xi;K)=S_g(\Lambda,x,\xi;K')
\Leftrightarrow\left\{
\begin{array}{l}
y\equiv 0 \\
\\
\theta\in SO(d)
\ .
\end{array}\right.
\ee
\par
It is now straightforward to translate these conclusions back to the
Lorentzian framework.
\par
As a simple corollary, one obtains that the action
$S_g$ in Eq.~(\ref{S_g}) is invariant under the left rigid
action of $ISO(d-1,1)$ iff $n=0$ or $n=d$.
\subsection{Equations of motion}
In order to study the equations of motion following from the
action $S_g$ it is more convenient to rewrite
\be
S_g=S^{(g)}+S^{(H)}+S^{(I)}
\ ,
\ee
where
\be
S^{(g)}=S_3+\int d^2\sigma\,\sum\limits_{r\not\in H}\,
\left[{\cal W}^{\,\alpha\beta}_{rpq}\,\partial_\alpha x^r\,
\partial_\beta\Lambda^{pq}+{1\over2}\,\sum\limits_{s\not\in H}\,
{\cal V}^{\,\alpha\beta}_{rs}\,\partial_\alpha x^r\,\partial_\beta x^s
\right]
\ ,
\ee
does not contain elements $x\in H$,
\be
S^{(H)}=\int d^2\sigma\,\sum\limits_{r\in H}\,
\left[{\cal W}^{\,\alpha\beta}_{rpq}\,\partial_\beta\Lambda^{pq}
+{1\over2}\,\sum\limits_{s\in H}\,{\cal V}^{\,\alpha\beta}_{rs}\,
(\partial_\beta x^s+\xi_\beta^s)\right]
\,(\partial_\alpha x^r+\xi_\alpha^r)
\ ,
\ee
contains only terms proportional to elements of $H$, and
\be
S^{(I)}=\int d^2\sigma\,\sum\limits_{r\in H,s\not\in H}\,
{\cal V}^{\,\alpha\beta}_{rs}\,(\partial_\alpha x^r+\xi_\alpha^r)
\,\partial_\beta x^s
\ ,
\ee
contains mixed terms.
\par
The equations of motion $\delta_x S_g=0$ together with
the transformation law for the gauge field in Eq.~(\ref{cond})
split the theory into two sectors:
\begin{enumerate}
\item
when $\delta x^i\not\in H$ one requires $\delta_x S^{(g)}=0$ and
obtains
\be
\partial_\alpha{\cal P}^{(g)\alpha}_{i}\equiv
\partial_\alpha\left({\cal P}^{(1g)\alpha}_i
+{\cal P}^{(2g)\alpha}_i\right)=0
\ ,\ \ \ \ i\not\in H
\ ,
\label{dpg=0}
\ee
so that the $d-\di(H)=d-n$ linear momentum currents
${\cal P}^{(g)\alpha}_i$ with $i\not\in H$ and
\be
\begin{array}{l}
{\cal P}^{(1g)\alpha}_{i}=\sum\limits_{j\not\in H}\,
{\cal V}^{\,\alpha\beta}_{ij}\,\partial_\beta x^j \\
{\cal P}_{i}^{(2g)\alpha}={\cal P}_{i}^{(2)\alpha}
\ ,
\end{array}
\ee
are still conserved;
\item
when $\delta x^i\in H$ one gets $\delta_x S^{(H)}=\delta_x S^{(I)}=0$
identically and thus the corresponding currents
\be
{\cal P}^{(H)\alpha}_i\equiv{\cal P}^{\alpha}_i-
{\cal P}^{(g)\alpha}_r=\sum\limits_{s\in H}\,
{\cal V}^{\,\alpha\beta}_{is}\,\partial_\beta x^s
\ ,\ \ \ \ i\in H
\label{Ph}
\ee
are not conserved.
\end{enumerate}
\par
From the variation $\delta_\Lambda S_g=0$ and Eq.~(\ref{cond})
one obtains an analogous splitting into three sectors:
\begin{enumerate}
\item
in the sector for which
$\delta \Lambda_{kj}=\Lambda_k^{\ i}\,\rho_{ij}$, $i,j\not\in H$
one has
\be
\partial_\alpha\left[{\cal J}^{\alpha}_{ij}
+2\,\sum\limits_{r\in H}\,
{\cal W}_{rij}^{\,\alpha\beta}\,\xi_\beta^r
+x_i\wedge \sum\limits_{r\in H}\,{\cal V}_{jr}^{\,\alpha\beta}\,
\xi_\beta^r\right]=0
\ \ \ \ \ i,j\not\in H
\ .
\label{djg=0}
\ee
Thus the angular momentum currents ${\cal J}^{\alpha}_{ij}$
with both indices $i,j\not\in H$ are no longer conserved
but couple to the gauge field;
\item
when $\delta \Lambda_{kj}=\Lambda_k^{\ i}\,\rho_{ij}$, $i,j\in H$
one obtains an analogous relation
\be
\partial_\alpha\left[{\cal S}_{ij}^\alpha
+2\,\sum\limits_{r\in H}\,{\cal W}^{\,\alpha\beta}_{rij}\,
\xi_\beta^r\right]=
\left({\cal P}^\alpha_i+\sum\limits_{r\in H}\,
{\cal V}_{ir}^{\,\alpha\beta}\,\xi_\beta^r\right)\wedge
\left(\partial_\alpha x_j+\xi_{\alpha j}\right)
\ \ \ \ i,j\in H\ ;
\label{H_mix}
\ee
in which only the spin ${\cal S}_{ij}$ appears on the L.H.S. because
now $\partial_\alpha{\cal P}^\alpha_i\not=0$, $i\in H$;
\item
finally, in the sector in which
$\delta \Lambda_{kj}=\Lambda_k^{\ i}\,\rho_{ij}$, $i\not\in H$, $j\in H$
one obtains the following non trivial equation
\begin{eqnarray}
&\partial_\alpha\left[{\cal J}^{\alpha}_{ij}
+2\,\sum\limits_{r\in H}\,
{\cal W}_{rij}^{\,\alpha\beta}\,\xi_\beta^r
+x_i\wedge \sum\limits_{r\in H}\,{\cal V}_{jr}^{\,\alpha\beta}\,
\xi_\beta^r\right]=&
\nonumber \\
&\left({\cal P}^\alpha_i+\sum\limits_{r\in H}\,
{\cal V}_{ir}^{\,\alpha\beta}\,\xi_\beta^r\right)\,
\left(\partial_\alpha x_j+\xi_{\alpha j}\right)
-\partial_\alpha x_i\,
\left({\cal P}^\alpha_j+\sum\limits_{r\in H}\,
{\cal V}_{jr}^{\,\alpha\beta}\,\xi_\beta^r\right)\, ,
\ \ i\not\in H,\, j\in H&
\ .
\label{H_sec}
\end{eqnarray}
which mixes terms from the two previous sectors.
\end{enumerate}
\par
To summarize, whenever one considers only quantities which
do not contain elements of $H$, the equations of motion for the linear
momenta look the same as the ungauged ones and amount again to linear
momentum conservation.
The angular momentum, instead, is no longer conserved, even
in that sector of the theory,
because of the presence of $S^{(H)}$ and $S^{(I)}$.
Further, due to these latter contributions to the action,
the constraints displayed in Eqs.~(\ref{H_mix}), (\ref{H_sec})
must also be satisfied, together with the equations of motion
for the gauge field, $\delta_\xi S_g=0$,
which we will study in the following Subsection.
\subsection{Eliminating $\xi$: the quadratic case}
Now that we have introduced the gauge field $\xi$,
we want to eliminate it from the action.
One way to achieve this goal is to integrate out $\xi$ first
in the path integral
\be
\int[d\Lambda]\,[dx\not\in{\cal H}]\,[dx\in{\cal H}]\,[d\xi]\,
e^{-S(\Lambda,x)-\Delta S(\Lambda,x,\xi)}
\equiv
\int[d\Lambda]\,[dx\not\in{\cal H}]\,\,e^{-S_{eff}(\Lambda,x)}
\ ,
\label{path}
\ee
where, from Eqs.~(\ref{ds1}), (\ref{ds2}) one has
\be
\Delta S\equiv\Delta S_1+\Delta S_2=\int d^2\sigma\,
\sum\limits_{s\in H}\,\left[{\cal P}^\alpha_s\,\xi^s_\alpha+
{1\over 2}\,\sum\limits_{r\in H}\,
{\cal V}^{\,\alpha\beta}_{rs}\,\xi^r_\alpha\,\xi^s_\beta
\right]
\ ,
\ee
with ${\cal P}^{\alpha}_s$ defined in Eq.~(\ref{dp=0}).
\par
When ${\cal V}^{\,\alpha\beta}_{rs}\not=0$, for some $r,s\in H$,
$\Delta S$ is quadratic in $\xi$ and the above mentioned integration
corresponds to solving the equations of motion for $\xi$,
namely $\delta_\xi\Delta S=0$,
and substituting back the result into the action.
We also notice that, since
${\cal V}^{\,\alpha\beta}=-{\cal V}^{\,\beta\alpha}$,
the condition ${\cal V}^{\,\alpha\beta}_{rs}\not=0$
implies that $\di(H)=n \ge 2$ and thus excludes
the ungauged case.
\par
It is easy to find that $\delta_\xi\Delta S=0$ implies
\be
{\cal P}^{\alpha}_s+\sum\limits_{r\in H}\,
{\cal V}^{\,\alpha\beta}_{sr}\,\xi^r_\beta=0
\ ,\ \ \ \ s\in H
\ .
\label{xi=P}
\ee
One can try to go further if we assume that
${\cal V}$ is invertible inside the $H$ sector
(this will put constraints on the constants $K^{(1)}$
and could also exclude part of the subgroup $SO(d-1,1)$)
and define ${\cal V}_{\alpha\beta}^{\,rs}$ such that
\be
\sum\limits_{a\in H}\,{\cal V}_{\alpha\gamma}^{\,ra}\,
{\cal V}^{\,\gamma\beta}_{as}=\delta_\alpha^\beta\,\delta^r_s
\ ,\ \ \ \ \forall\,r,s\in H
\ .
\ee
Then Eq.~(\ref{xi=P}) above can be inverted to give
\be
\xi^r_\alpha=-\sum\limits_{s\in H}\,
{\cal V}_{\alpha\beta}^{\,rs}\,{\cal P}^{\beta}_s
\ ,\ \ \ \ r\in H
\ ,
\label{xi=}
\ee
and the correction to the (ungauged) action becomes
\be
\Delta S=-{1\over 2}\,\int d^2\sigma\,
\sum\limits_{r,s\in H}\,{\cal P}^{\alpha}_r\,
{\cal V}_{\alpha\beta}^{\,rs}\,{\cal P}^{\beta}_s
\ .
\label{corr}
\ee
By simply expanding the linear momentum current ${\cal P}$
according to Eq.~(\ref{Ph}) one then easily proves
that $\Delta S$ cancels out every dependence on $x\in H$ from
the (ungauged) action, since
\begin{eqnarray}
&S_1(x\not\in H,x\in H)-{1\over 2}\,
\int d^2\sigma\,\sum\limits_{r,s\in H}\,{\cal P}^{(1H)\alpha}_r\,
{\cal V}^{rs}_{\alpha\beta}\,\left({\cal P}^{(1H)\beta}_s
+2\,{\cal P}^{(1g)\beta}_s\right)=
S_1(x\not\in H)&
\nonumber \\
&S_2(x\not\in H, x\in H)-
\int d^2\sigma\,\sum\limits_{r,s\in H}\,{\cal P}^{(1H)\alpha}_r\,
{\cal V}^{rs}_{\alpha\beta}\,{\cal P}^{(2g)\beta}_s=
S_2(x\not\in H)&
\ .
\end{eqnarray}
The remaining terms in Eq.~(\ref{corr}) lead to corrections
to the three (ungauged) contributions of the action
according to the following scheme
\begin{eqnarray}
S^q_{1eff}(x\not\in H)&=&S_1(x\not\in H)-{1\over 2}\,\int
d^2\sigma\,\sum\limits_{r,s\in H}\,{\cal P}^{(1g)\alpha}_r\,
{\cal V}^{rs}_{\alpha\beta}\,{\cal P}^{(1g)\beta}_s
\nonumber \\
S^q_{2eff}(x\not\in H)&=&S_2(x\not\in H)-
\int d^2\sigma\,\sum\limits_{r,s\in H}\,{\cal P}^{(1g)\alpha}_r\,
{\cal V}^{rs}_{\alpha\beta}\,{\cal P}^{(2g)\beta}_s
\nonumber \\
S^q_{3eff}&=&S_3-
{1\over 2}\,\int d^2\sigma\,\sum\limits_{r,s\in H}\,
{\cal P}^{(2g)\alpha}_r\,{\cal V}^{rs}_{\alpha\beta}\,
{\cal P}^{(2g)\beta}_s
\ .
\label{cor_q}
\end{eqnarray}
This defines a total effective action
\be
S^q_{eff}(\Lambda,x\not\in H,K)=S^q_{1eff}+S^q_{2eff}+S^q_{3eff}
\ ,
\ee
which differs from the one obtained by simply setting the terms
containing $x\in H$ to zero.
This is due to the presence of {\em cross terms} of the kind
discussed in the Introduction.
\par
We observe that the number of degrees of freedom in the
effective action $S^q_{eff}$ is at most
$N\equiv\di(ISO(d-1,1))-\di(H)=d\,(d+1)/2-n$, $n<d$,
and that the only case (with $n\ge 2$)
which is invariant under the left rigid action of $ISO(d-1,1)$,
namely $n=d$, is given by $S^q_{eff}=S^q_{3eff}$
and has $N=d\,(d-1)/2$.
\par
Further, although the gauged action $S_g$ is not well behaved
under the action of any local subgroup of the Lorentz group
$SO(d-1,1)$, as we have shown in Subsection 3.1, it turns out that the
effective theory can be made invariant
under the local action of the group $SO(N-D,D)$ in $N$
space-time dimensions,
$0\le 2\,D\le N$ (the precise signature must be computed for each case
explicitly).
This is not difficult to prove.
Let us consider the $N$-dimensional vector $\bar x\equiv(x,t)$
whose components are the $d-n$ translations $x\not\in H$ and the
$d(d-1)/2$ independent parameters $t$ of the Lorentz group $SO(d-1,1)$.
We notice that the effective action $S^q_{eff}$ written in terms of
$\bar x$ is of the form
\be
S_{eff}={1\over 2}\,\int d^2\sigma\,\bar M_{ab}^{\alpha\beta}(t)\,
\partial_\alpha \bar x^a\,\partial_\beta \bar x^b
\ ,
\label{S_eff1}
\ee
where
\be
\bar M_{ab}=\bar M_{ab}^{(1)}+\bar M_{ab}^{(2)}+\bar M_{ab}^{(3)}
\ ,
\ee
is an $N\times N$ matrix whose elements depend on (some of)
the coordinates $t$ (and the constants $K$) only.
It can be written in the following block form:
\begin{eqnarray}
\begin{array}{lcr}
\bar M^{(1)}=\left[\begin{array}{cc}
\epsilon^{\alpha\beta}\,B_{xx}^{(1)}\ & 0 \\
0 & 0
\end{array}\right]
 &
\bar M^{(3)}=\left[\begin{array}{cc}
 0\ &  0 \\
0 & g^{\alpha\beta}\,G_{tt}^{(3)}+
\epsilon^{\alpha\beta}\,B_{tt}^{(3)}
\end{array}\right]
\end{array}
\ ,
\end{eqnarray}
where the symmetric matrix $G_{tt}$ is $d(d-1)/2 \times d(d-1)/2$
dimensional and comes from the corrections to $S_3$ sketched
in the last of Eqs.~(\ref{cor_q}), and
the antisymmetric matrices $B_{xx}$ and $B_{tt}$
are respectively $(d-n)\times(d-n)$ and
$d\,(d-1)/2\times d\,(d-1)/2$ dimensional;
\be
\bar M^{(2)}=\left[\begin{array}{cc}
0 & (g^{\alpha\beta}+2\,\epsilon^{\alpha\beta})\,U_{xt}^{(2)}
\\
(g^{\alpha\beta}-2\,\epsilon^{\alpha\beta})\,U_{xt}^{(2)\, T}
& 0
\end{array}\right]
\ ,
\ee
where the matrix $U_{xt}$ is $(d-n)\times d\,(d-1)/2$ dimensional.
Thus $S_{eff}$ can also be written
\begin{equation}
S_{eff}={1\over 2}\,\int d^2\sigma\,\left[g^{\alpha\beta}\,
\bar G_{ab}\,\partial_\alpha\bar x^a\,\partial_\beta\bar x^b+
\epsilon^{\alpha\beta}\,
\bar B_{ab}\,\partial_\alpha\bar x^a\,\partial_\beta\bar x^b
\right]
\ ,
\label{S_eff}
\end{equation}
where $\bar G$ is the symmetric part of $\bar M$,
\be
\bar G=\left[\begin{array}{cc}
0 & U_{xt}^{(2)}     \\
U_{xt}^{(2)\,T} & G_{tt}^{(3)}
\end{array}\right]
\ ,
\label{G}
\ee
and $\bar B$ is the antisymmetric part,
\be
\bar B=\left[\begin{array}{cc}
B_{xx}^{(1)} &  2\,U_{xt}^{(2)}              \\
-2\,U_{xt}^{(2)\, T} & B_{tt}^{(3)}
\end{array}\right]
\ .
\label{B}
\ee
\par
Now $S_{eff}$ in Eq.~(\ref{S_eff}) is in the form of a $\sigma$-model
which describes a bosonic string moving in a
(generally curved) $(N-D)+D$-dimensional background parameterized
by the coordinates $\bar x$.
Such a model, with the addition of a possible dilaton field (see
Section~\ref{T-d}), is invariant under the local action of the
group $SO(N-D,D)$ as claimed, although the matter source
which generates it might be unphysical.
\par
The symmetric matrix $\bar G$ is not zero iff $S_2\not\equiv 0$
and $g^{\alpha\beta}\not\equiv 0$.
It plays the role of the metric tensor in $N$ dimensions,
its signature thus determining the number of time-like coordinates $D$.
The matrix  $\bar B$ is the antisymmetric potential
of the axion field
$\tilde H_{abc}\equiv\partial_a \bar B_{bc}+\partial_c \bar B_{ab}
+\partial_b \bar B_{ca}$.
\subsection{Eliminating $\xi$: the linear case}
\label{linear}
When ${\cal V}_{rs}^{\,\alpha\beta}=0$, $\forall\,r,s\in H$,
one has that $\Delta S$ is linear in $\xi$.
This case includes both the ungauged action (for which $n=0$)
and all the 1-dimensional subgroups $H\sim\real$.
\par
Upon taking $\delta_\xi \Delta S=0$ one would get
\be
{\cal P}^{(g)\alpha}_s=0
\ ,\ \ \ \ s\in H
\ ,
\ee
but one now is not allowed to substitute this result back
into the action.
However, we observe that, since the action $S_g$
does not contain a kinetic term for $\xi$,
one can force $\xi$ to be a {\em pure gauge\/},
\be
\xi^r_\alpha=-\partial_\alpha x^r\ ,
\ \ \ \ r\in H
\ ,
\label{gauge}
\ee
by adding a Lagrange multiplier term,
\be
\int d^2\sigma\,
\lambda_r\,\epsilon^{\alpha\beta}\,\partial_\alpha\xi_\beta^r
\ ,
\ee
to the exponent in Eq.~(\ref{path}) and then integrating out
$\lambda$.
\par
The relation in Eq.~(\ref{gauge}) reduces Eq.~(\ref{djg=0})
to the conservation of the angular momentum current
${\cal J}^{(g)}_{ij}$, $i,j\not\in H$, defined by
\be
{\cal J}^{(g)}_{ij}\equiv {\cal J}^{(1g)}_{ij}+{\cal J}^{(2g)}_{ij}
+{\cal J}^{(3g)}_{ij}
\ ,
\ee
where
\begin{eqnarray}
{\cal J}_{ij}^{(1g)\alpha}&=&{\cal L}_{ij}^{(1g)\alpha}=
x_i\wedge{\cal P}_{j}^{(1g)\alpha}
\nonumber \\
{\cal J}_{ij}^{(2g)\alpha}&=&{\cal L}_{ij}^{(2g)\alpha}
+{\cal S}_{ij}^{(2g)\alpha}
\ ,\ \ \ \
\left\{\begin{array}{l}
{\cal L}_{ij}^{(2g)\alpha}={\cal L}_{ij}^{(2)\alpha}  \\
 \\
{\cal S}_{ij}^{(2g)\alpha}={\cal S}^{(2)\alpha}_{ij}-
{\cal S}^{(2H)\alpha}_{ij}=
2\sum\limits_{r\not\in H}\,{\cal W}_{rij}^{\,\alpha\beta}\,
\partial_\beta x^r
\end{array}\right.
\nonumber \\
{\cal J}_{ij}^{(3g)\alpha}&=&{\cal J}_{ij}^{(3)\alpha}
\ .
\end{eqnarray}
It thus follows that one is left with only the equations of motion
for $(\Lambda,x\not\in H)$,
$\delta_x S^{pg}_{eff}=\delta_\Lambda S^{pg}_{eff}=0$
derived by varying the effective action obtained this time by
setting to zero terms for which $x\in H$ in the ungauged action,
\be
S^{pg}_{eff}(\Lambda,x\not\in H;K)
=S^{pg}_{1eff}+S^{pg}_{2eff}+S^{pg}_{3eff}
\ ,
\label{S^pg_eff}
\ee
where
\begin{eqnarray}
S_{1eff}^{pg}&=&
{1\over 2}\,\int d^2\sigma\,\sum\limits_{r,s\not\in H}\,
{\cal V}_{rs}^{\,\alpha\beta}\,
\partial_\alpha x^r\,\partial_\beta x^s
\nonumber  \\
S_{2eff}^{pg}&=&
\int d^2\sigma\,\sum\limits_{r\not\in H}\,
{\cal W}_{rsk}^{\,\alpha\beta}\,
\partial_\alpha x^r\,\partial_\beta\Lambda^{sk}
\\
S_{3eff}^{pg}&=&S_3
\ ,
\nonumber
\label{S_123eff}
\end{eqnarray}
and the sums $\sum_{r,s,\ldots\not\in H}$ run only over the indices
corresponding to the translations not included in $H$.
\par
The effective action $S^{pg}_{eff}$ expressed in terms of the coordinates
$\bar x$ is again of the same form given in Eq.~(\ref{S_eff}) but,
since $S^{pg}_{3eff}=S_3$,
the matrix $G_{tt}^{(3)} \equiv 0$ in Eq.~(\ref{G}) and one finds that
the metric tensor is represented by an $N\times N$ square matrix of
dimension $N=d\,(d+1)/2-n$,
\be
\bar G=\left[\begin{array}{cc}
0 & U_{xt}^{(2)}     \\
U_{xt}^{(2)\,T} & 0
\end{array}\right]
\ ,
\label{Gpg}
\ee
where the matrix $U_{xt}^{(2)}$ is again $(d-n)\times d\,(d-1)/2$
dimensional.
\par
It is easy to prove that each matrix of the block form above
is degenerate and admits
\be
N_d=\left|{d\,(d-1)\over 2}-(d-n)\right|
\ee
zero eigenvalues.
This implies that the dimension of the non-degenerate subspace
is only
\be
N-N_d=\left\{\begin{array}{ll}
2\,(d-n)\ , &\ \ \ \ {d\,(d-1)\over 2}-(d-n)\ge 0 \\
\\
d\,(d-1)\ , &\ \ \ \ {d\,(d-1)\over 2}-(d-n) < 0
\ ,
\end{array}\right.
\ee
so that, if one gauges the whole $d$-dimensional
translation group ($n=d$) the effective theory $S^{pg}_{eff}=S_3$
has no metric structure and becomes spatially empty.
Only internal (originally interpreted as spin) degrees of freedom
(the ones contained in $S_3$)
survive, and the present reduction scheme looks quite singular.
\par
Contrary to the metric tensor, the antisymmetric matrix $\bar B$
is, in general, non-singular due to the presence of $B_{tt}$
in Eq.~(\ref{B}) and one obtains an axion field potential
in a $N$-dimensional space.
This makes the overall picture quite pathological, unless one regards
the extra (degenerate) dimensions as pure parameters of the theory.
In so doing, one ends up with a $N_d$-parameter family of $N-N_d$
dimensional $\sigma$-models.
\par
However, there are exceptional cases in which the metric $\bar G$ is
not singular.
This happens when $U_{xt}^{(2)}$ is square, or $N_d=0$,
for which one obtains that the matrix $\bar G$ is even
dimensional and necessarily has nonzero eigenvalues $\pm G_i$
with $i=1,\ldots,N/2$.
There are only two such cases:
\begin{enumerate}
\item
$d=2$, $n=1$ with $N=2$;
and
\item
$d=3$, $n=0$ with $N=6$,
\par\noindent
\end{enumerate}
which we will examine in Sections \ref{d=2} and respectively
\ref{iso21}, where we prove that this last case reduces
to the one studied in Ref.~\cite{ch}.
\setcounter{equation}{0}
\section{T-duality}
\label{T-d}
Since both $\bar G$ and $\bar B$ in Eq.~(\ref{S_eff})
can depend at most on the $d\,(d-1)/2$ Lorentz parameters $t$,
the action $S^{q}_{eff}$ clearly displays at least $d-n$ target space
isometries corresponding to the $d-n$ (translational) coordinates $x$
in Eq.~(\ref{S_eff1}).
This fact can be used to introduce transformations that define T-dual
spaces (see Ref.~\cite{giveon} and references therein for a general
exposition).
\par
The linear case, for which we have just
shown that the metric is degenerate, is a special case because the
effective dimension of the space-time is lower than $N$.
In this case one could eliminate $N_d$ dimensions by simply
diagonalizing $\bar G$ and one would end up with
$(N-N_d)-d\,(d-1)/2$ isometries.
However this process would mix $x$ and $t$ coordinates, thus
making less transparent the identification of the isometric
directions.
For this reason in the present Section we neglect the degeneracies
of the metric structure in $S^{pg}_{eff}$ and work in the full
$N$-dimensional space.
This allows us to develop a formal treatment which is valid for both
$S^{pg}_{eff}$ and $S^{q}_{eff}$.
One has only to remember that in the following sections
$G_{tt}=0$ for $S_{eff}=S^{pg}_{eff}$.
\subsection{Dual Actions}
\par
To begin with, we want to show some of the general features
of our model which are related to the isometry being (possibly) more
than $1$-dimensional in the case in which none of the
Lorentz parameters $t$
corresponds to an isometry of the action.
\par
First one doubles the $d-n$ coordinates $x_1\equiv x$
by adding an equal number of new coordinates $x_2$.
Then one introduces a parent action in the new
$N+(d-n)=d\,(d+3)/2-2\,n$
dimensional space with coordinates $(x_1,x_2,t)$,
\be
S_{N+d-n}(x_1,x_2,t)=S_{xt}+S_t
\ ,
\label{S_double}
\ee
where
\begin{eqnarray}
S_{xt}&=&{1\over 2}\,\int d^2\sigma\,\left[D^{\alpha\beta}_{ab}\,
\left(\partial_\alpha x_1^a\,\partial_\beta x_1^b
+\partial_\alpha x_2^a\,\partial_\beta x_2^b\right)
+2\,\Sigma_{ab}^{\alpha\beta}\,\partial_\alpha x_1^a\,\partial_\beta x_2^b
\right.
\nonumber \\
&&\phantom{{1\over 2}\,\int d^2\sigma\,[}\left.
+2\,N_{ai}^{\alpha\beta}\,\left(\partial_\alpha x_1^a\,\partial_\beta t^i
+\partial_\alpha x_2^a\,\partial_\beta t^i\right)\right]
\nonumber \\
S_t&=&{1\over 2}\,\int d^2\sigma\,\left[T^{\alpha\beta}_{ij}\,
\partial_\alpha t^i\,\partial_\beta t^j
+\Phi\,R^{(2)}\right]
\ ,
\label{S_eff+}
\end{eqnarray}
and we have also defined
\begin{eqnarray}
D_{ab}^{\alpha\beta}&\equiv& \epsilon^{\alpha\beta}\,B_{xx,ab}
\nonumber \\
N_{ai}^{\alpha\beta}&\equiv&
\left(g^{\alpha\beta}+2\,\epsilon^{\alpha\beta}\right)\,U_{xt,ai}
\nonumber \\
\Sigma_{ab}^{\alpha\beta}&\equiv& g^{\alpha\beta}\,\Sigma_{ab}^S
+\epsilon^{\alpha\beta}\,\Sigma^A_{ab}
\nonumber \\
T^{\alpha\beta}_{ij}&\equiv&g^{\alpha\beta}\,G_{tt,ij}
+\epsilon^{\alpha\beta}\,B_{tt,ij}
\ ,
\label{DNST}
\end{eqnarray}
with $a,b=1,\dots,d-n$ and $i,j=1,\dots,d\,(d-1)/2$.
Here the matrices $\Sigma_{ab}^S(t)$ (symmetric), $\Sigma_{ab}^A(t)$
(antisymmetric) and the dilaton field $\Phi(t)$,
which couples to the world-sheet scalar curvature
$R^{(2)}$, have been introduced so that the world-sheet
theory described by $S_{N+d-n}$ is conformal.
It is well known that a dilaton $\Phi$ must be included
whenever the metric is a non-vacuum solution of the Einstein equations.
Therefore it must satisfy \cite{witten,gsw}
\be
\nabla_r\nabla_s\Phi=R_{rs}
\ ,
\label{Phi}
\ee
where $\nabla_r$ is the covariant derivative in the target background
space-time and $R_{rs}$ is the Ricci tensor.
When the dimension of space-time is not $26$,
one must also add the central term $(N+d-n-26)/3$.
\par
Suppose now we impose the condition that the coordinates $x_2$ are
periodic
\be
x_2\equiv x_2+2\,\pi
\ .
\ee
In so doing, we are actually compactifying
coordinates which correspond to one of the two copies
of the (non-compact) translational
parameters of the Poincar\'e group in the gauged action in
Eq.~(\ref{S_g}).
We are then left with only $(d-n)+d-1$ non-compact
($x_1$ and boost parameters) and $d\,(d-3)/2+1$ compact
(rotation angles) coordinates.
The action $S_{N+d-n}$ is now manifestly invariant under the
$U(1)^{d-n}$ affine symmetries acting on $x_2$ which are generated
by the $d-n$ currents
\be
J_{2\,a}^\alpha= D_{ab}^{\alpha\beta}\,\partial_\beta x_2^b
+\Sigma_{ab}^{\alpha\beta}\,\partial_\beta x_1^b
+N_{ai}^{\alpha\beta}\,\partial_\beta t^b
\ .
\ee
One can gauge this ($d-n$)-dimensional symmetry
by minimal coupling, introducing a gauge field $A_2$ such that
\be
\partial_\alpha x_2\to\partial_\alpha x_2+A_{2\,\alpha}
\ ,
\ee
and one gets a gauged action given by
\be
S_{N+d-n}^g(x_1,x_2,t,A_2)=S_{N+d-n}+
\int d^2\sigma\,A_{2\,\alpha}^a\,\left[ J_{2\,a}^\alpha
+{1\over 2}\,D_{ab}^{\alpha\beta}\,A_{2\,\beta}^b\right]
\ .
\label{S_g2}
\ee
\par
When $A_2$ is a pure gauge, $A_{2\,\alpha}=-\partial_\alpha x_2$,
one is led back to the effective action in $N$ dimensions
$S_{eff}$ displayed in Eq.~(\ref{S_eff})
(plus the possible dilaton field) which does not contain $x_2$.
\par
The equations of motion $\delta_{A_2} S_{N+d-n}^g=0$ imply that
\be
A_{2\alpha}^a=-D_{\alpha\beta}^{ab}\,J_{2b}^\beta
\ ,
\ee
where we have assumed that $D_{ab}^{\alpha\beta}$ is invertible and
we have defined $D_{\alpha\beta}^{ab}$ such that
\be
D_{\alpha\gamma}^{ac}\,D_{cb}^{\gamma\beta}=\delta_\alpha^\beta\,
\delta_b^a
\ .
\ee
On substituting the solution for $A_2$ into Eq.~(\ref{S_g2})
one obtains an effective action
\be
S_N\equiv S_{N+d-n}-{1\over 2}\,\int d^2\sigma
J_{2a}^\alpha\,D_{\alpha\beta}^{ab}\,J_{2b}^\beta
\ ,
\ee
where $S_N$ does not depend on $x_2$ and is given by
\begin{eqnarray}
S_N(x_1,t)&=&{1\over 2}\,\int d^2\sigma\,\left[
\left(T^{\alpha\beta}_{ij}
+{1\over 2}\,N_{ai}^{\alpha\gamma}\,D_{\gamma\lambda}^{ab}\,
N_{bj}^{\lambda\beta}\right)\,
\partial_\alpha t^i\,\partial_\beta t^j
+\Phi\,R^{(2)}\right]
\nonumber \\
&&
+{1\over 2}\,\int d^2\sigma\,\left[
N_{ai}^{\alpha\beta}
+{1\over 2}\,N_{bi}^{\alpha\gamma}\,D_{\gamma\lambda}^{bc}\,
\Sigma_{ca}^{\lambda\beta}
+{1\over 2}\,\Sigma_{ac}^{\alpha\gamma}\,D_{\gamma\lambda}^{cd}\,
N_{di}^{\lambda\beta}\right]\,\partial_\alpha x_1^a\,\partial_\beta t^i
\ .
\label{S_N1}
\end{eqnarray}
The conclusion is that $S_N$ can be obtained from
$S_{N+d-n}$ by setting $\partial x_2=0$ and changing
\begin{eqnarray}
T^{\alpha\beta}_{ij}&\to&T^{\alpha\beta}_{ij}
-{1\over 2}\,N_{ai}^{\alpha\gamma}\,D_{\gamma\lambda}^{ab}\,
N_{bj}^{\lambda\beta}
\nonumber \\
N_{ai}^{\alpha\beta}&\to&N_{ai}^{\alpha\beta}
-{1\over 2}\,N_{bi}^{\alpha\gamma}\,D_{\gamma\lambda}^{bc}\,
\Sigma_{ca}^{\lambda\beta}
+{1\over 2}\,\Sigma_{ab}^{\alpha\gamma}\,D_{\gamma\lambda}^{bc}\,
N_{ci}^{\lambda\beta}
\ ,
\end{eqnarray}
which are the T-duality transformations in the present context.
\par
On compactifying and gauging $x_1$ instead of $x_2$,
one would obtain the same action as the one given in Eq.~(\ref{S_g2})
with only an interchange of the labels for the two sets of $d-n$
isometric coordinates $x_1$ and $x_2$.
\par
But suppose we define new coordinates which are linear
combinations of $x_1$ and $x_2$,
thus rotating and scaling the directions in which we
compactify in the $(x_1,x_2)$ subspace.
This would change the quantities in Eq.~(\ref{DNST}).
\par
One example is given by
\be
\left\{\begin{array}{l}
x\equiv (x_1+x_2)/2 \\
\\
\tilde x\equiv (x_2-x_1)/2
\ .
\end{array}\right.
\label{x=x}
\ee
We now impose the periodicity requirement on the coordinates
$\tilde x$,
\be
\tilde x=\tilde x+2\,\pi
\ .
\ee
The contribution $S_{xt}$ in the action $S_{N+d-n}$ becomes
\begin{eqnarray}
S_{xt}&=&{1\over 2}\,\int d^2\sigma\,\left[
D^{\alpha\beta}_{ab}\,
\partial_\alpha x^a\,\partial_\beta x^b
+\tilde D^{\alpha\beta}_{ab}\,
\partial_\alpha\tilde x^a\,\partial_\beta\tilde x^b
\right.
\nonumber \\
&&\phantom{{1\over 2}\,\int d^2\sigma\,[}\left.
+2\,N_{ai}^{\alpha\beta}\,\partial_\alpha x^a\,\partial_\beta t^i
-2\,\Sigma^{\alpha\beta}_{ab}\,\partial_\alpha x^a\,\partial_\beta\tilde
x^b \right]
\ ,
\label{Sxt}
\end{eqnarray}
where
\begin{eqnarray}
D_{ab}^{\alpha\beta}&\equiv&2\,g^{\alpha\beta}\,\Sigma_{ab}^S
+2\,\epsilon^{\alpha\beta}\,\left(B_{xx,ab}+\Sigma^A_{ab}\right)
\nonumber \\
\tilde D_{ab}^{\alpha\beta}&\equiv&-2\,g^{\alpha\beta}\,\Sigma_{ab}^S
+2\,\epsilon^{\alpha\beta}\,\left(B_{xx,ab}-\Sigma^A_{ab}\right)
\nonumber \\
N_{ai}^{\alpha\beta}&\equiv&
\left(g^{\alpha\beta}+2\,\epsilon^{\alpha\beta}\right)\,U_{xt,ai}
\nonumber \\
\Sigma_{ab}^{\alpha\beta}&\equiv&\epsilon^{\alpha\beta}\,\Sigma^A_{ab}
\ .
\end{eqnarray}
The whole action $S_{N+d-n}$ is now manifestly invariant under the
$U(1)^{d-n}$ affine symmetries acting on $\tilde x^a$
which are generated by the currents
\be
\tilde J_a^\alpha=\tilde D^{\alpha\beta}_{ab}\,\partial_\beta\tilde x^b
-2\,\Sigma_{ab}^{\alpha\beta}\,\partial_\beta x^b
\ .
\ee
As before, one can gauge this ($d-n$)-dimensional symmetry
by minimal coupling introducing a gauge field $\tilde A$ such that
\be
\partial_\alpha\tilde x\to\partial_\alpha\tilde x+\tilde A_\alpha
\ ,
\ee
and one gets a gauged action given by
\be
S_{N+d-n}^g(x,\tilde x,\tilde A)=S_{N+d-n}+
{1\over 2}\,\int d^2\sigma\,\tilde A_\alpha^a\,\left[\tilde J_a^\alpha
+\tilde D^{\alpha\beta}_{ab}\,\tilde A_\beta^b\right]
\ ,
\ee
which, on using the equations of motion for the gauge field,
$\delta_{\tilde A} S_{N+d-n}^g=0$,
\be
\tilde A_{\alpha}^a=-\tilde D_{\alpha\beta}^{ab}\,\tilde J_{b}^\beta
\ ,
\ee
becomes the new effective action in $N$ dimensions,
\be
S_N(x,t)=S_t+{1\over 2}\,\int d^2\sigma\,\left[
{D'}^{\alpha\beta}_{ab}\,\partial_\alpha x^a\,\partial_\beta x^b
+2\,N_{ai}^{\alpha\beta}\,\partial_\alpha x^a\,\partial_\beta t^i
\right]
\ ,
\ee
where the T-duality transformation is
\be
D\to D'=D-4\,\Sigma\,\tilde D^{-1}\,\Sigma
\ .
\ee
Now $S_{N}$ does not contain $\tilde x$ and is again different from
the action $S_{eff}$ in Eq.~(\ref{S_eff}).
\par
If one instead compactifies and gauges $x$, the corresponding currents
are given by
\be
J_a^\alpha=D^{\alpha\beta}_{ab}\,\partial_\beta x^b
+N^{\alpha\beta}_{ai}\,\partial_\beta t^i
-2\,\Sigma_{ab}^{\alpha\beta}\,\partial_\beta\tilde x^b
\ .
\ee
One obtains an effective action which depends on $\tilde x$ alone
and can be obtained by setting $\partial x=0$ everywhere
in Eq.~(\ref{Sxt}) above, with
\begin{eqnarray}
T&\to&T-N\,D^{-1}\,N
\nonumber \\
\tilde D&\to&\tilde D-4\,\Sigma\,D^{-1}\,\Sigma
\nonumber \\
N&\to&N+2\,N\,D^{-1}\,\Sigma+2\,\Sigma\,D^{-1}\,N
\ .
\end{eqnarray}
\par
One could then conclude that a different linear combination
of the form given in Eq.~(\ref{can}) for our toy model,
\be
\left[\begin{array}{c}
x \\
\tilde x
\end{array}\right]
\equiv
\left[\begin{array}{cc}
\Theta_{11} & \Theta_{12} \\
\Theta_{21} & \Theta_{22}
\end{array}\right]
\,
\left[\begin{array}{c}
x_1 \\
x_2
\end{array}\right]
\ ,
\label{gen_T}
\ee
in place of Eq.~(\ref{x=x}) would lead to a different
$N$-dimensional $\sigma$-model and all of them are T-duals
of $S_{eff}$ in Eq.~(\ref{S_eff}).
However, in order to prove that the space-times corresponding to
different
choices of the matrix $\Theta$ are physically different, one should
be able to compare the scalar curvatures and the other invariant
quantities of General Relativity, including the axion field.
\subsection{Dualizing one isometric coordinate}
In the following Sections we will consider examples in which
we want to compute the dual of the action with respect to
one isometric coordinate.
Thus we now specialize the results obtained so far to this simpler
case.
\par
From the definition in Eq.~(\ref{DNST}) it follows that,
when the set of isometric coordinates $x_1$ we gauge is given by
$x_1^1$ only, $D^{\alpha\beta}_{11}=0$
and no quadratic term in $A_1$ nor $A_2$ will ever appear.
This implies the gauge field must be a pure gauge and we
obtain the action $S_{eff}$ we started from before we doubled the
coordinate $x_1^1$.
\par
In order to get a non trivial answer, we have to perform a
transformation of the type displayed in Eq.~(\ref{gen_T}),
{\em e.g.} the one in Eq.~(\ref{x=x}).
Then we have
\begin{eqnarray}
D^{\alpha\beta}_{11}&=&2\,g^{\alpha\beta}\,\Sigma_{11}^S
\nonumber \\
\tilde D^{\alpha\beta}_{11}&=&-2\,g^{\alpha\beta}\,\Sigma_{11}^S
\ ,
\end{eqnarray}
which we can expect to be in general different from zero.
If we gauge $\tilde x^1$, we obtain a generating current
\be
\tilde J^\alpha_1=-2\,g^{\alpha\beta}\,\Sigma_{11}^S\,
\left(\partial_\beta \tilde x^1+\partial_\beta x^1\right)
+\sum\limits_{b>1}\,\left[\tilde D^{\alpha\beta}_{1b}\,
\partial_\beta\tilde x^b-2\,\Sigma_{1b}^{\alpha\beta}\,
\partial_\beta x^b\right]
\ ,
\ee
for the corresponding $U(1)$ symmetry, in which we singled out
the first term in the sum.
\par
If the terms with $b>1$ vanish in the expression for the current
above, the ungauged action (before one doubles $x^1$) reads
\be
S_{eff}=S_t+\int d^2\sigma\,N^{\alpha\beta}_{1i}\,\partial_\alpha x^1\,
\partial_\beta t^i
\ ,
\label{1-dim}
\ee
while the gauged action becomes
\be
S_{N+d-n}^g(x,\tilde x,\tilde A^1)=S_{N+d-n}-
\int d^2\sigma\,\tilde A_\alpha^1\,g^{\alpha\beta}\,\Sigma^S_{11}\,
\left[\partial_\beta \tilde x^1+\partial_\beta x^1
+\tilde A_\beta^1\right]
\ ,
\ee
and one obtains
\be
S_N(x,t)=S_t+\int d^2\sigma\,\left[g^{\alpha\beta}\,\Sigma^S_{11}\,
\partial_\alpha x^1\,\partial_\beta x^1+
N^{\alpha\beta}_{1i}\,\partial_\alpha x^1\,\partial_\beta t^i\right]
\ ,
\ee
where the metric tensor $\bar G$ has now acquired the new component
$G_{xx,11}=2\,\Sigma^S_{11}$.
\setcounter{equation}{0}
\section{The lowest dimensional case: $ISO(1,1)$}
\label{d=2}
We now consider the  $(d=2)$-dimensional case for which the
algebra is simple enough to allow one to carry the computation
to the end.
\par
Every element $\Lambda\in SO(1,1)$ can be written as function
of the only boost parameter $t\in\real$,
\be
\begin{array}{lcr}
\Lambda^i_{\ j}=\left[\begin{array}{cc}
\cosh t&\sinh t \\
\sinh t&\cosh t
\end{array}\right]
&\ \ \ \ \ \ \ &
\Lambda^{\ i}_j=\left[\begin{array}{cc}
\cosh t&-\sinh t \\
-\sinh t&\cosh t
\end{array}\right]
\ .
\end{array}
\ee
The relevant 1-forms in Eq.~(\ref{gdg}) become
\be
\begin{array}{lr}
V^i=\left[\begin{array}{l}
\cosh t\,dx^1-\sinh t\,dx^2 \\
\cosh t\,dx^2-\sinh t\,dx^1
\end{array}\right]
\ \ &\ \
W^{ij}=\left[\begin{array}{cc}
0&+1\\
-1&0
\end{array}\right]\,dt
\ .
\end{array}
\ee
It is then easy to find that
\begin{eqnarray}
S_1&=&S_3=0
\nonumber \\
S_2&=&\int d^2\sigma\,g^{\alpha\beta}\,
\left[\left(K_2\,\sinh t-K_1\,\cosh t\right)\,
\partial_\alpha x^1\,\partial_\beta t \right.
\nonumber \\
&&\left.\phantom{2\,\int d^2\sigma\,\epsilon^{\alpha\beta}\,[}
+\left(K_1\,\sinh t-K_2\,\cosh t\right)\,
\partial_\alpha x^2\,\partial_\beta t\right]
\ ,
\label{S(1,1)}
\end{eqnarray}
where total derivatives are discarded as usual and
$(K_1,K_2)$ are the only relevant independent constants satisfying
Eq.~(\ref{K_cond}) in $d=2$.
\par
The (conserved) linear momentum currents related to $S_2$ are given by
\begin{eqnarray}
{\cal P}^\alpha_1&=&g^{\alpha\beta}\,
\left(K_1\,\cosh t-K_2\,\sinh t\right)\,\partial_\beta t
\nonumber \\
{\cal P}^\alpha_2&=&g^{\alpha\beta}\,
\left(K_2\,\cosh t-K_1\,\sinh t\right)\,\partial_\beta t
\ .
\end{eqnarray}
Upon varying the parameter $t$ one finds that the following quantity is
also conserved
\begin{eqnarray}
{\cal J}^\alpha&=&{\cal P}^\alpha_1\,x^2- {\cal P}^\alpha_2\,x^1
+g^{\alpha\beta}\,
\left(K_1\,\cosh t-K_2\,\sinh t\right)\,\partial_\beta x^1
\nonumber \\
&&+
g^{\alpha\beta}\,
\left(K_2\,\cosh t-K_1\,\sinh t\right)\,\partial_\beta x^2
\ .
\end{eqnarray}
It is quite obvious that, since only the term proportional to
$g^{\alpha\beta}$ survives, no axion field will appear
in the resulting $\sigma$-models.
Further, one can gauge only 1-dimensional subgroups,
since eliminating both $x^1$ and $x^2$ leads to $S^{pg}_{eff}=0$.
\subsection{Gauging a 1-dimensional subgroup}
Since $S_2$ is linear in both $x^1$ and $x^2$,
if we gauge a 1-dimensional subgroup, {\em e.g.\/} the one corresponding
to $x^2$, we obtain that the gauge field $A^2$ is a pure gauge,
$A^2=-\partial x^2$.
This is actually the first of the two exceptional cases listed
in Section~\ref{linear}.
\par
The conserved currents which survive are given by
${\cal P}_1$ and ${\cal J}(x^2=0)$ above.
We then define
\be
\left\{\begin{array}{l}
X\equiv x^1+t \\
\\
T\equiv x^1-t
\ ,
\end{array}\right.
\ee
and we obtain
\begin{eqnarray}
S^{pg}_{2eff}&=&{1\over 2}\,\int d^2\sigma\,
g^{\alpha\beta}\,f(X-T)\,\left(
\partial_\alpha X\,\partial_\beta X-
\partial_\alpha T\,\partial_\beta T  \right)
\ ,
\end{eqnarray}
where $f(t)\equiv K_2\,\sinh t-K_1\,\cosh t$.
The diagonal form of the metric tensor is thus given by
\be
\bar G=\left[\begin{array}{cc}
f & 0 \\
0 & -f
\end{array}\right]
\ ,
\label{G2}
\ee
which becomes singular (both components vanish)
for $f(t_s)=0\ \Leftrightarrow \ \tanh t_s=K_1/K_2$, $K_2\not=0$.
\par
The curvature of space-time and the Ricci tensor are zero
everywhere, so $\bar G$ above represents a vacuum solution.
The singularity $t=t_s$ is a light-like volume singularity whose location
depends on the ratio of the constants $K_1/K_2$.
For example, when $K_1=0$ and $K_2=1$, one finds that
$\bar G$ becomes singular along the light-cone $X=T$.
\subsection{T-dual form}
The action $S^{pg}_{eff}$ is of the form given in Eq.~(\ref{1-dim})
with $S_t=\int d^2\sigma\,\Phi\,R^{(2)}$ and
$N^{\alpha\beta}_{11}=2\,g^{\alpha\beta}\,f(t)$.
If we introduce a coordinate $x_2^1$ and define $x$ and
$\tilde x$ according to Eq.~(\ref{x=x}),
we can then dualize with respect to the coordinate $\tilde x$ and
obtain a new metric tensor whose diagonal form is given by
\be
\bar G=\left[\begin{array}{cc}
\Sigma^S_{11}-\sqrt{(\Sigma^S_{11})^2+f^2(t)}
& 0 \\
0 &
\Sigma^S_{11}+\sqrt{(\Sigma^S_{11})^2+f^2(t)}
\end{array}\right]
\ .
\label{G2T}
\ee
Regardless of the explicit form of $\Sigma^S_{11}=\Sigma^S_{11}(t)$,
$\det(\bar G)=f^2$ and one obtains the same volume singularity for
$f(t_s)=0$.
The scalar curvature is again zero everywhere.
\par
The causal structure determined by $\bar G$ in Eq.~(\ref{G2T})
is different from the one given by the metric tensor in Eq.~(\ref{G2}),
since in the latter case one has an overall change of sign when
going through $f=0$, however in the former this cannot happen.
\par
Further, according to Eq.~(\ref{Phi}), since the Ricci tensor is
still zero, the presence of a non-vanishing $\Sigma^S_{11}(t)$
does not affect the dilaton.
\setcounter{equation}{0}
\section{$S=S_2$ in $d=3$ dimensions}
\label{iso21}
The second exceptional case listed in Section~\ref{linear}
has $d=3$ and $n=0$.
Since we are mainly interested in the metric structure of the
effective theory, we only consider $S=S_2(K)$ according to the
general form given in Eq.~(\ref{Gpg}).
\par
First we prove that, when $g^{\alpha\beta}=\eta^{\alpha\beta}$
there is actually only one such action,
namely the one with $K^{(2)}_{ijk}=\epsilon_{ijk}$,
and one recovers the model previously studied in Ref.~\cite{ch}.
In fact, every matrix $K^{(2)}_{ijk}$ with the symmetry properties
displayed in Eq.~(\ref{K_cond}) can be written
\be
K^{(2)}_{ijk}=\sum\limits_{l=0}^2\,
A^{(l)}_i\,\Sigma^{(l)}_{jk}
\ ,
\ee
where
\be
\begin{array}{ccc}
\Sigma^{(0)}=
\left[\begin{array}{ccc}
0 & 1 & 0 \\
-1 & 0 & 0 \\
0 & 0 & 0
\end{array}\right]\ ,\ \ &
\Sigma^{(1)}=
\left[\begin{array}{ccc}
0 & 0 & -1 \\
0 & 0 & 0 \\
1 & 0 & 0
\end{array}\right]\ ,\ \ &
\Sigma^{(2)}=
\left[\begin{array}{ccc}
0 & 0 & 0 \\
0 & 0 & 1 \\
0 & -1 & 0
\end{array}\right]\ ,\
\end{array}
\ee
and $A_i^{(l)}$ are nine arbitrary constants such that
$\det(A_i^{(l)})\not=0$.
The action $S_2$ then can be written
\begin{eqnarray}
S_{(2,1)}(\Lambda,y;K)&=&
{1\over 2}\,\int d^2\sigma\,
A^{(l)}_i\,\Sigma^{(l)}_{jk}\,\partial_-y^i
(\partial_+\Lambda\,\Lambda^{-1})^{jk}
\nonumber \\
&=&{1\over 2}\,\int d^2\sigma\,A^{(l)}_i\partial_-y^i
\,\Tr\left[\Sigma^{(l)}\,
(\partial_+\Lambda\,\Lambda^{-1})\right]
\ ,
\end{eqnarray}
where $y\in\real^3$ and $\sigma^\pm$ have been defined in
Eq.~(\ref{light}).
The trace in the integrand above can now be evaluated assuming
a specific parameterization of the Lorentz group $SO(2,1)$.
As in Ref.~\cite{ch}, we write any matrix $\Lambda^i_{\ j}$
as a product of two rotations (of angles $\alpha$ and $\gamma$)
and a boost ($\beta$),
\be
\Lambda=
\left[\begin{array}{ccc}
1 & 0 & 0 \\
0 & \cos\alpha & -\sin\alpha \\
0 & \sin\alpha & \cos\alpha
\end{array}\right]\,
\left[\begin{array}{ccc}
\cosh\beta & 0 & \sinh\beta \\
0 & 1 & 0 \\
\sinh\beta & 0 & \cosh\beta
\end{array}\right]\,
\left[\begin{array}{ccc}
1 & 0 & 0 \\
0 & \cos\gamma & -\sin\gamma \\
0 & \sin\gamma & \cos\gamma
\end{array}\right]
\ ,
\ee
and we obtain
\begin{eqnarray}
\Tr\left[\Sigma^{(0)}\,
(\partial_+\Lambda\,\Lambda^{-1})\right]
&\equiv&
{\cal P}^0_+=\partial_+t^0+\cosh t^1\,\partial_+ t^2 \\
\nonumber  \\
\Tr\left[\Sigma^{(1)}\,
(\partial_+\Lambda\,\Lambda^{-1})\right]
&\equiv&
{\cal P}^1_+=\cos t^0\,\partial_+ t^1
+\sin t^0\,\sinh t^1\,\partial_+t^2  \\
\nonumber \\
\Tr\left[\Sigma^{(2)}\,
(\partial_+\Lambda\,\Lambda^{-1})\right]
&\equiv&
{\cal P}^2_+=\sin t^0\,\partial_+ t^1
-\cos t^0\,\sinh t^1\,\partial_+t^2
\ .
\label{P21}
\end{eqnarray}
Finally, on defining
\be
x^i\equiv\sum\limits_{j=0}^2\,A^{(i)}_j\,y^j
\ ,
\label{x=y}
\ee
one gets
\be
S_{(2,1)}(\Lambda,y;K)=
\int d^2\sigma\,{\cal P}_+^k\,\partial_-x_k
=S_{(2,1)}(\Lambda,x;\epsilon_{ijk})
\ ,
\label{2+1}
\ee
as claimed.
\par
This result is very peculiar and follows from the fact that
there are $d^2\,(d-1)/2$ independent elements in $K^{(2)}_{ijk}$.
This number is equal to the square of the space-time dimension
for $d=3$ and one can thus use all these constants to build
the linear combination given in Eq.~(\ref{x=y}).
In general (for $d>3$) one has
\be
d^2\,(d-1)/2>d^2
\ ,
\ee
and one can not eliminate in this way $d^2\,(d-3)/2$ elements of
$K^{(2)}_{ijk}$.
\par
The three linear momentum currents in Eq.~(\ref{P21})
define a metric tensor $\bar G$ in 6 dimensions of the
form given in Eq.~(\ref{Gpg}) with
\be
U_{xt}=\left[\begin{array}{ccc}
1  &  0  &  \cosh t^1 \\
0  &  \cos t^0  &  \sin t^0\,\sinh t^1  \\
0  &  \sin t^0  &  -\cos t^0\,\sinh t^1
\end{array}\right]
\ .
\ee
\par
The Ricci tensor computed from the metric $\bar G$ has the following
non-zero components,
\begin{eqnarray}
\bar R_{t^0 t^0}&=&-1
\nonumber\\
\bar R_{t^0 t^2}&=&-\cosh(t^1)
\nonumber\\
\bar R_{t^1 t^1}&=&1
\nonumber\\
\bar R_{t^2 t^2}&=&-1
\ ,
\end{eqnarray}
and its trace $\bar R=0$.
The subspace $t^1=0$ is a volume singularity, since
\be
\det(\bar G)=f^2(t^1)
\ ,
\label{g}
\ee
with $f\equiv\sinh t^1$.
\par
We now analyze two degenerate cases following from $S_{(2,1)}$.
\subsection{Gauging a 1-dimensional translation}
We notice that $S_{(2,1)}$ is already invariant under the following
{\em semi\/}-local action of the Poincar\'e group:
\be
g\to h_{_L}(\sigma^+)\,g\,h_{_R}^{-1}(\sigma^-)
\ ,
\ee
where $h_{_{L/R}}=(\theta_{_{L/R}},y_{_{L/R}})\ \in\ ISO(2,1)$.
However, it is not invariant under the {\em fully\/}
local action of any subgroup $H$ of $ISO(2,1)$ given by
$g\to h_{_L}\,g\,h_{_R}^{-1}=
\left(\theta_{_L}\,\Lambda\,\theta^{-1}_{_R},
-\theta_{_L}\,\Lambda\,\theta^{-1}_{_R}\,y_{_R}+
\theta_{_L}\,x+y_{_L}\right)$,
where $h_{_{L/R}}=h_{_{L/R}}(\sigma^-,\sigma^+)=
(\theta_{_{L/R}},y_{_{L/R}})\in H$,
due to the dependence of $h_{_L}$ on $\sigma^-$
and of $h_{_R}$ on $\sigma^+$.
To promote $H$ to a gauge symmetry of the action we introduce again
the gauge field $A_\pm=\left(\omega_\pm,\xi_\pm\right)\in iso(2,1)$,
and the covariant derivatives $D_\pm g=\partial_\pm g+A_\pm$.
The requirement that $H$ acts invariantly,
$\delta g=h_{_L}\,g=(0,y_{_L})$,
leads to $h_{_{L/R}}=(0,y_{_{L/R}}\in\real^n)$, $n\le 3$,
so that
\be
\left\{\begin{array}{lr}
\omega_\pm=\xi_+ \equiv 0 &\\
 \\
\xi_-^k \equiv 0 & k\not\in H
\ .
\end{array}\right.
\ee
\par
If we gauge the 1-dimensional subgroup corresponding to $x^0$,
the gauged action $S_g(x,t,\xi^0_-)$ is linear in $\xi_-^0$
and after we eliminate $t^0$ as an irrelevant parameter, we obtain the
effective action \cite{ch}
\begin{eqnarray}
S^{pg}_{eff}(x^1,x^2,t^1,t^2)&=&\int d^2\sigma\,\left[
-\partial_+ t^1\,\partial_- x^1+
\sinh t^1\,\partial_+ t^2\,\partial_- x^2\right]
\nonumber \\
&=&\int d^2\sigma\,\left(\eta^{\alpha\beta}
+\epsilon^{\alpha\beta}\right)\,\left[
\partial_\alpha t^1\,\partial_\beta x^1
-f\,\partial_\alpha t^2\,\partial_\beta x^2\right]
\ ,
\label{S_4}
\end{eqnarray}
where again $f=\sinh t^1$.
The metric tensor is $N-N_d=5-1=4$-dimensional and
is given by
\be
\bar G=\left[\begin{array}{cccc}
0 & 0 & 1 & 0 \\
0 & 0 & 0 & -f \\
1 & 0 & 0 & 0 \\
0 & -f & 0 & 0
\end{array}\right]
\ .
\label{G21}
\ee
The axion field potential in this non-degenerate 4-dimensional
subspace is given by
\be
\bar B=2\,\left[\begin{array}{cccc}
0 & 0 & 1 & 0 \\
0 & 0 & 0 & -f \\
-1 & 0 & 0 & 0 \\
0 & f & 0 & 0
\end{array}\right]
\ ,
\label{B21}
\ee
its field strength $\bar H$ having $\bar H_{124}=-2\,\cosh t^1$ as
the only non-zero component.
\par
The Ricci tensor in this frame of reference has one
non-vanishing component,
\be
\bar R_{t^1 t^1}={1\over 2}\,{\sinh^2 t^1-1\over\sinh^2 t^1}
\ ,
\ee
and the scalar curvature $\bar R$ is zero.
\par
The signature of the metric is $2+2$ and never changes,
as can be inferred by noting that the determinant of
$\bar G$ is again given by the expression in Eq.~(\ref{g}),
and the eigenvalues of $\bar G$ are given by $(\pm 1,\pm f)$.
This corrects an erroneous statement in the last part of
Ref.~\cite{ch}.
\subsection{T-dual form}
Both the metric tensor $\bar G$ and the axion field $\bar B$ given
above depend only on $t^1$, so that in this case we have a Lorentz
parameter ($t^2$) which is irrelevant, as are the two translational
parameters $x^1$ and $x^2$.
\par
The dual effective theory obtained upon introducing
$x$ and $\tilde x$ as given in Eq.~(\ref{x=x}) with $x^1_1\equiv x^1$
contains the same axion field potential $\bar B$ given in
Eq.~(\ref{B21}) above, but the metric tensor acquires a new component
\be
\bar G_{xx}=2\,\Sigma(t^1)
\ .
\ee
This extra term generates new non-zero components for the Ricci
tensor,
\begin{eqnarray}
\bar R_{x x}&=&2\,\Sigma\,\dot\Sigma\,{\sqrt{f^2+1}\over f}
+\ddot\Sigma
\nonumber\\
\bar R_{x t^1}&=&-\ddot\Sigma
-\dot\Sigma\,{f\over\sqrt{f^2+1}}
\nonumber\\
\bar R_{x^2 t^2}&=&\Sigma\,f
+\dot\Sigma\,\sqrt{f^2+1}
\ ,
\end{eqnarray}
where $\dot\Sigma\equiv\partial_{t^1}\Sigma$,
and a non-vanishing scalar curvature,
\be
\bar R=-3\,\Sigma-2\,\ddot\Sigma+{\Sigma\over f^2}
-4\,\sqrt{f^2+1}\,{\dot\Sigma\over f}
\ .
\ee
However the determinant of the metric is still given by Eq.~(\ref{g}),
while the eigenvalues of $\bar G$ become
$(\Sigma\pm\sqrt{\Sigma^2+1},\pm f)$.
\subsection{Gauging a 2-dimensional translation}
Upon gauging a 2-dimensional subgroup one obtains an effective action
in $N-N_d=4-2=2$ dimensions of the same type as the one in
Eq.~(\ref{S_eff}), but with $\bar G$ a constant symmetric matrix with
signature 1+1 and $\bar B$ a constant antisymmetric matrix.
\par
On dualizing with respect to the unique translational coordinate
which is left, one obtains a metric tensor whose diagonal form
is
\be
\bar G=\left[\begin{array}{cc}
\Sigma-\sqrt{\Sigma^2+1} & 0 \\
0 & \Sigma+\sqrt{\Sigma^2+1}
\end{array}\right]
\ ,
\ee
in which there are no singularities regardless of the specific
form of $\Sigma=\Sigma(t)$.
\section{Conclusions}
In this paper we have examined a new class of $\sigma$-models, which are
generated by gauging a subgroup of the Poincar\'e group $ISO(d-1,1)$
which acts invariantly from the left.
The fact that this group is a noncompact, semi-direct product group
differentiates our coset model from all such models studied heretofore.
There are several intriguing results in this investigation.
Our starting point is a model with values in $ISO(d-1,1)$,
which describes spinning strings in flat $(d-1)+1$ dimensions.
After promoting a translation subgroup to a gauge symmetry,
however, the resulting action describes spinless strings moving
in curved space-times and interacting with an axion field.
If the effective action is obtained from a pure gauge field,
the resulting metric tensor is in general degenerate.
The degeneracy is equal to the difference between the number
of relevant (Lorentz) coordinates and the number of isometries.
Finally, the effective actions inherently possess T-duals.
\bigskip
\par
\begin{center}
{\bf Acknowledgements}:
\end{center}
\par
We would like to thank A. Stern for many useful discussions.
This work was supported in part by the U.S. Department of
Energy under Grant No. DE-FG02-96ER40967.
\bigskip

\end{document}